\documentclass[A4,aps,twocolumn,showpacs,groupedaddress]{revtex4}

\usepackage{amssymb}
\usepackage{amsmath}
\usepackage{dcolumn}
\usepackage{bm}
\usepackage{graphicx}

\begin{document}

\title{Feedback controlled electro\-migration in four-terminal nano-junctions}

\author{ZhengMing Wu}
\author{M. Steinacher}
\author{R. Huber}
\author{M. Calame}
\author{S. J. van der Molen}
\author{C. Sch{\"o}nenberger}

\email{christian.schoenenberger@unibas.ch}

\affiliation{Institut f{\"u}r Physik, Universit{\"a}t Basel, Klingelbergstr.~82, CH-4056 Basel, Switzerland }

\begin{abstract}

We have developed a fast, yet highly reproducible method to
fabricate metallic electrodes with nanometer separation using
electro\-migration (EM). We employ four-terminal instead of
two-terminal devices in combination with an analog feedback to
maintain the voltage $U$ \emph{over} the junction constant. After
the initialization phase (\mbox{$U \alt 0.2$\,V}), during which
the temperature $T$ increases by \mbox{$80-150$\,$^o$C}, EM sets
in shrinking the wire locally. This quickly leads to a transition
from the diffusive to a quasi-ballistic regime (\mbox{$0.2$\,V
$\alt U \alt 0.6$\,V}). At the end of this second regime, a gap
forms (\mbox{$U \agt 0.6$\,V}). Remarkably, controlled electro\-migration
is still possible in the quasi-ballistic regime.
\end{abstract}



\maketitle

Single-molecule electronics has been the focus of substantial
worldwide research~\cite{Review_Single_Mol_Elec}. Direct measurement
of electron flow through a single molecule promises a better
understanding of the electron transfer processes in molecules. To
measure single molecules, small metallic junctions
with nano-sized gaps are needed in between which molecules can then
anchored and electrically measured.

Various methods have been developed to define and measure such
molecular junctions~\cite{Techniques}. Among these,
electro\-migration (EM) induced nano-gaps have successfully been
employed for a broad range of molecules, revealing various
transport phenomena~\cite{Park2000,JPark2002,Liang2002}.
EM-junctions have the advantage that gates with a decent
gate-to-molecule coupling can be fabricated~\cite{Liang2002}.
However, the junction formation is a slow process and prone to instabilities.
In addition, nano-particles can form during the EM-process through which
electric transport may occur subsequently~\cite{Houck2005,Sordan2005,Heersche2006,Strachan2006}.
Refined EM processes are therefore highly desirable.
In this article we introduce a new technique that employs a fast
analog electronic feedback to accurately control the voltage over
the junction during the EM process.

Electro\-migration is the directed migration of atoms caused by a
large electric current density. EM proceeds by momentum transfer
from electrons to atoms and requires sufficient atom mobility to
occur. The latter increases at higher temperatures, so that local
Joule heating is an important parameter in addition to current
density~\cite{Black1969}. The formation of an EM nano-gap starts with
the lithographic definition of a metallic wire with a constriction (junction),
where the EM process will be effective. EM narrows the junction down,
until a gap forms and the process self-terminates.
In such lithographically defined wires, the
bonding pads are far away from the constriction, yielding long leads
with comparatively large lead resistances $R_L$. Typically, $R_L$ is
much larger than the resistance of the junction $R_J$ (Fig.~1,
inset). Although a voltage $U_0$ is applied, the junction is
effectively current-biased through $R_L$. Consequently, as EM starts
shrinking the junction and $R_J$ increases,
the power dissipated on the junction grows proportional to $R_J$,
which may cause a thermal run-away, destroying the junctions.
This instability appears in the  $I-U_0$ characteristics (see Fig.~1)
along branch B, which is multi-valued. In the shaded region,
the junction can rapidly be destroyed, if it switches, for example,
at point \textbf{p} to the open state well above the breaking point \textbf{e}.
Because this happens at much larger power dissipation than would be
the case at point \textbf{e}, the junction is ``burnt'' off.
In order not to destroy the junction, one therefore has to ensure that
the process follows branch B. This can be done manually, or better
by software control~\cite{Strachan2005,Esen2005}.
This approach is however quite slow, as $U_0$ needs to be set back
and slowly ramped up repetitively.
It would be much better to remove the destructive region altogether.
Point \textbf{s} occurs at larger $U_0$ values
than point \textbf{e}, because $R_L >> R_J$. Hence, designing devices
with low lead resistances relaxes the problem~\cite{Trouwborst2006}.
We eliminate the lead resistances by defining \emph{four} terminals to each
junction and applying a novel and fast electronic feedback scheme.

\begin{figure}[!htb]
\begin{center}
\includegraphics[width=85mm]{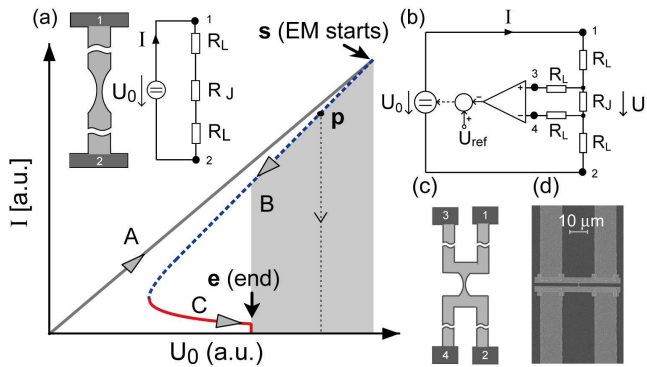}
\end{center}
\caption{Schematic characteristic of the current $I$ versus bias
  voltage $U_0$ during software controlled 2-terminal
  electro\-migration (EM)~\cite{Strachan2005,Esen2005}. EM starts at point
  \textbf{s} and the junction breaks open at point \textbf{e}. Insets:
  (a) schematics of the constriction forming the junction with attached
  leads and its equivalent circuit diagram. $R_L$, $R_J$ are the lead
  and junction resistances, respectively.
  (b) The electric circuit for 4-terminal EM and (c)
  an SEM micrograph of an actual device. The voltage drop $U$ over the
  junction is maintained constant and equal to a preset reference value U$_{ref}$
  by the feedback system.
} \label{figure1}
\end{figure}

The principle of our EM procedure is illustrated in Fig.~1b. The
voltage drop over the junction $U$ is controlled by a custom-made
feedback voltage source. The four terminals are defined by two
symmetric pairs of contacts, a left and
a right pair (Fig.~1c-d). On one pair, the bias voltage $U_0$ is
supplied, while on the other the voltage drop $U$ over the junction is
simultaneously measured.
Regardless of the actual value of the junction resistance
$R_J$, the feedback voltage source maintains $U$ constant. This
removes the thermal instability, because if $U=const$ while $R_J$
evolves to larger values due to EM, the power over
the junction decreases. A nano-gap is formed by ramping up $U_{ref}=U$ until the
junction switches to a high-ohmic state with \mbox{$R_J >
100$\,k$\Omega$} at \mbox{$U\backsimeq 0.4\dots 0.6$\,V}. This is
typically performed during a few minutes, but can be done faster~\cite{comment_speed}
or slower with no observable difference. Because we would like to
characterize the junction during the evolution of EM, we do not ramp
$U$ continuously but in a square-wave pattern. This is illustrated in
Fig.~2c. We measure $R_J(U)$ at voltage $U$ (arrows) and
subsequently switch to $U\simeq 0$ to measure the instantaneous
linear-response resistance of the junction $R^{0}_J(t)$ with the aid
of a small voltage modulation (lock-in technique). Although
$R^{0}_J(t)$ is measured at $U\simeq 0$, we plot it as a function of
$U$, enabling the comparison of $R_J(U)$ with $R^{0}_J$. The whole
process is performed at room temperature under ambient conditions.

Our devices are fabricated with two sequential lift-off processes on oxidized \mbox{($400$\,nm)} Si
substrates using \mbox{$50$\,nm} Au together with a \mbox{$5$\,nm} thick Ti adhesion layer, where
the latter is absent in the narrow junction region. The typical size of a fabricated Au
constriction is \mbox{$200$\,nm} in length and  \mbox{$100$\,nm} in width. The resistance of the
junction $R_J$ is around \mbox{$3-10$\,$\Omega$} at room temperature whereas the overall resistance
$R=R_J+2 R_L$ typically amounts to as much as \mbox{$250$\,$\Omega$}.

\begin{figure}[!htb]
\begin{center}
\includegraphics[width=85mm]{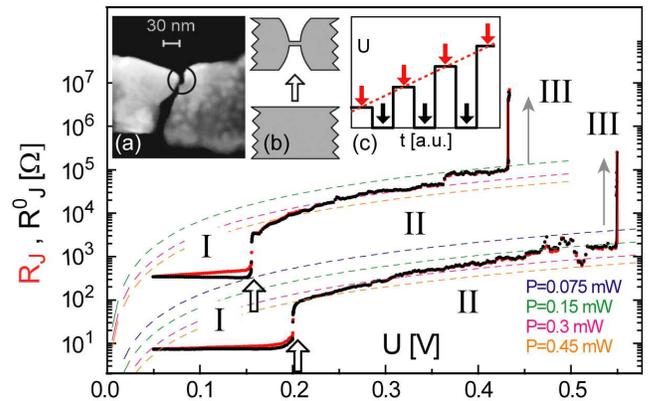}
\end{center}
\caption{Evolution of the junction resistance as function of junction voltage $U$ during 4-terminal
  EM at room temperature for two devices. The upper curves are shifted by two orders of magnitude
  for clarity. The voltage $U$ is ramped up in a square-wave pattern as illustrated in (c).
  After applying $U\neq 0$ during a short period and measuring the junction resistance $R_J(U)$ (red
  arrows), $U$ is switched back to zero for a similar period of time during which the equilibrium
  resistance $R^0_J$ is measured (black arrows). A typical cycle lasts \mbox{$0.1$\,s}. The graph
  shows both $R_J(U)$ and $R^0_J$ during the whole process until the junction  switches open into the
  tunneling regime with \mbox{$R_J \agt 100$\,k$\Omega$}. The dashed curves are drawn as reference
  lines and correspond to constant power values, i.e. $P=U^2/R_J=const$. $P$ decreases from bottom
  \mbox{($0.45$\,mW)} to top \mbox{($75$\,$\mu$W)}.
  (a) an SEM micrograph of a typical junction after feedback EM and
  (b) an illustration of the slit formation after EM started.
} \label{figure2}
\end{figure}

Two representative graphs of the evolution of the junction
resistance $R_J(U)$ (red) and the corresponding equilibrium
resistance $R^0_J$ (black), measured while ramping up the junction
voltage $U$, are shown in Fig.~2. Three regimes ({\rm I-III}) can be
discerned: in regime {\rm I}, the constant equilibrium resistance
$R^0_J$ shows that geometrically nothing happens. The sudden, but controlled
increase in $R_J$ (arrow) at $U=0.15\dots 0.2$\,V signals the
transition to regime {\rm II}. Because $R^0_J$ has increased by
typically one order of magnitude, the cross-section of the junction
has consequently been decreased. $R_J$ grows
steadily with increasing junction voltage $U$, showing that EM is
active. There is a second sudden jump occurring typically between
\mbox{$U=0.4$\,V} and \mbox{$0.6$\,V}. In this transition to regime {\rm
III}, $R_J$ grows from \mbox{$\approx 1$\,k$\Omega$} to large
values \mbox{$> 100$\,k$\Omega$}. Due to the large current drop, EM
stops at this point leaving the junction `open'. In regime III, a
gap has been formed and the device shows tunneling
behavior. We indeed measure non-linear $I(U)$ characteristics that
follow the expected Simmons-law~\cite{Simmons1963} quite well in this
regime. Fig.~2a  shows that EM tends
to form slits that are typically smaller than \mbox{$30$\,nm} in width.
Within these slits, there is a small part (indicated by a circle),
which is even narrower. It is here that the gap is formed.
More than $20$ samples have been processed with this feedback method and in
$18$ of them, EM proceeded smoothly in the manner described before.

There are two remarkable features visible in Fig.~2. In the first
place, $R_J$ significantly differs from $R^0_J$ in regime {\rm I},
whereas in regime {\rm II}, $R_J$ and $R^0_J$ are almost equal.  In
the second place, the transition from regime {\rm I} to {\rm II}, although
appearing as a step, is gradual and rapidly flattens off again.
Below, we argue that both these features point to a transition
from the diffusive regime (regime {\rm I}) to a
`quasi-ballistic' regime (regime {\rm II}).

We first discuss regime {\rm I}, which is diffusive given
the size of our constriction. Upon increasing the voltage $U$,
EM does not start immediately, as confirmed by a constant $R^0_J$.
As $U$ increases further, the current density in the
constriction increases. This leads to a higher local temperature,
as witnessed by the increase of $R_J(U)$
with respect to $R^0_J$. The temperature increase yields a strong
rise in the atomic mobility~~\cite{Sordan2005,Trouwborst2006,Esen2005,Lambert2003}
and at a certain voltage, typically \mbox{$U \alt 0.2$\,V}, EM becomes considerable and
the constriction starts to narrow down.

To estimate the local temperature close to the onset of EM,
the difference between the junction resistances $R_J(U)$ and $R^0_J$ can
be used~\cite{Trouwborst2006,Stahlmecke2005}. It can be related to a temperature
difference $\Delta T$ alone if two conditions hold: i) the geometry
does not change in between subsequent measurements of $R_J(U)$ and
$R^0_J$ (small time delay); ii) the inelastic scattering length
$l_{in}$ is much smaller than the length of the junction $L$
(diffusive regime). In this case, we may write
$R_J(U)=R^0_J\cdot(1+\alpha\Delta T)$, where $\alpha$ is a constant
which has to be measured independently. To do so, the resistivity
$\rho$ of a thin gold film with equal thickness
was measured as a function of temperature $T$ in the
vicinity of \mbox{$T=25$\,$^o$C}. $\rho(T)$ increases with $T$
according to $(\rho(T)-\rho_0)/\rho_0=0.9\cdot\Delta T$. Here,
$\rho_0$ denotes the resistivity at \mbox{$T=25$\,$^o$C}.

\begin{figure}[!htb]
\begin{center}
\includegraphics[width=85mm]{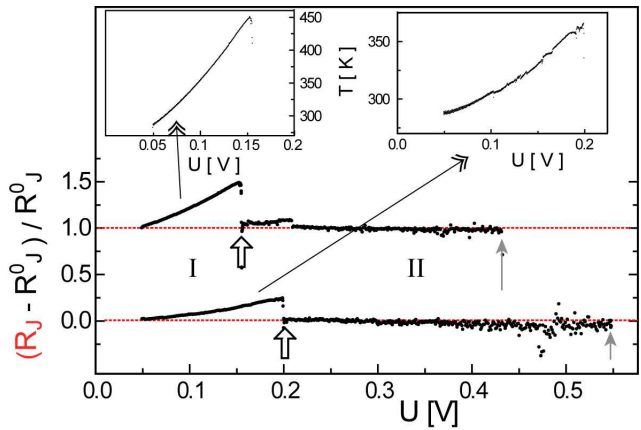}
\end{center}
\caption{The difference of the junction resistance R (U)  and R$_{0}$ (U=0). The two curves
  correspond to the same process as in Fig.~2. The up curve is shifted by 1 for clarity. In the
  insets are the development of local temperature in the junction before EM begins. The left inset is
  for the upper curve and the right inset for the lower curve. The procedure is carried out at room
  temperature.
} \label{figure3}
\end{figure}

In Fig.~3, we present $(R_J-R^0_J)/R^0_J$ as a function of the
applied junction voltage $U$. The data correspond to the same two
samples that gave rise to the measurements in Fig.~2. That $R_J$
increases above $R^0_J$ in regime {\rm I}, as we have emphasized before,
can now be seen much clearer. The corresponding
temperature increase is shown in the upper two insets.
We see that $\Delta T$ reaches maximum values of
\mbox{$180$\,$^o$C} and \mbox{$90$\,$^o$C} respectively,
proving that a substantial temperature increase is
required for EM to be initiated. This has been anticipated
before~\cite{Trouwborst2006,Strachan2005,Esen2005} and is confirmed here.

Once regime {\rm II} is entered, the difference between $R_J$ and
$R^0_J$ is surprisingly small. However, EM still takes place as
evidenced by an increase in both $R_J$ and $R^0_J$. At first sight,
this suggests that EM proceeds close to room temperature. While this
conclusion is tempting, it rests on the assumption that the
inelastic scattering length $l_{in}$ remains shorter than the
effective junction length $L$ in regime {\rm II} as well. However,
after entering regime {\rm II}, the junction has narrowed and
effectively shortened. In fact, SEM images such as the one in
Fig.~2a indicate that the size of the slit is smaller than
\mbox{$30$\,nm}, a value close to the electron mean-free path ~\cite{comment}.
This would then imply that a cross-over in the transport regime has taken place,
from diffusive (`viscous' to be more precise) with $l_{in}<<L$,
to quasi-ballistic, with $l_{in} \simeq L$. We can then understand
why our `thermometer' ceases to work in regime {\rm II},
because the resistance depends
only slightly on temperature in the quasi-ballistic regime.

This picture becomes even more plausible, if we closely look at the
data of Fig.~3 in regime {\rm II}. The junction resistance $R_J(U)$
even slightly decreases compared to its
equilibrium value $R^0_J$ as EM evolves, as if the temperature would
decrease. This effect is very weak in the upper data set, but
remarkably pronounced in the lower. It has been observed in the
majority of electro\-migrated devices. This lowering can be
understood if we assume that the current-voltage ($I-U$)
characteristics becomes non-linear. This is the case when only a few
scattering centers remain along the length of the
junction. In the extreme case of a single scattering center (a
tunnel barrier), $I(U)$ is not linear and increases stronger than
linear above a characteristic energy scale, determining the strength
of the scattering center. This again supports our conclusion that
the effective junction length becomes shorter than the
$l_{in}$ in regime {\rm II}, turning viscous electron motion into a
quasi-ballistic one. This picture explains why the fast transition
from regime {\rm I} to regime {\rm II} flattens off (see Fig.~2)
and proceeds smoothly and well controlled down to the atomic scale.
It does so because scattering is greatly reduced.

It may be considered surprising that EM proceeds at all in the
quasi-ballistic regime. Although the number of scattering events
decreases and slowing down EM in regime {\rm II}, it implies that
there is still enough scattering at the constriction to induce narrowing.
To remove the last few atoms in the constriction, one needs to
increase the bias by almost a factor of $3$ to finally create a gap.
During this process, the total dissipation is not
constant, as conjectured by two groups~\cite{Houck2005,Strachan2005}, but decreases (Fig.~2).
To our knowledge, little work has been done on EM-induced narrowing of
quasi-ballistic constrictions~\cite{Ralls1989_and_Holweg1992}.
Understanding this paradoxical situation, will be advantageous for our full
understanding of EM. This may prove beneficial for semiconductor
industry, which uses thinner and thinner interconnects between
devices.

This work has been supported by the Swiss National Center (NCCR) on
``Nanoscale Science'', the Swiss National Science Foundation, and the
European Science Foundation through the Eurocore program on
Self-Organized Nanostructures (SONS). S.J.v.d.M. acknowledges The
Netherlands Organization for Scientific Research, NWO (``Talent
stipendium'').


\begin{thebibliography}{7}
\bibitem{comment_speed} %
The response time of this feedback system is better than \mbox{$0.5$\,$\mu$s}.
\bibitem{comment} 
We estimate a mean-free path of \mbox{$20$\,nm}
from the measured sheet resistance \mbox{$R_{\square}\sim 1$\,$\Omega$}
of similar Au films using Drude's formula.
\end{thebibliography}
\end{document}